Review Article

# Electron-Phonon Coupling on the Surface of Topological Insulators

Xuetao Zhu[1], Colin Howard[2], Jiandong Guo[1]
and Michael El-Batanouny[2]

*1. Beijing National Laboratory for Condensed Matter Physics and Institute of Physics, Chinese Academy of Sciences, Beijing, China*

*2. Department of Physics, Boston University, Boston, MA, USA*

## Abstract

Topological insulators (TIs) are materials that have a bulk electronic band gap like an ordinary insulator but have protected conducting states on their surface. One of the most interesting properties of TIs is their spin helicity, whereby the spin is locked normal to the wave vector of the surface electronic state. The topological surface states should be very stable in TIs, since these spin-textured surface states are robust against spin-independent backscattering. Scattering from defects and other lattice imperfections is possible provided the spin is not completely flipped. However, the quality of TI crystals can be controlled by careful growth, whereas phonons will exist in even the most perfect crystals. Consequently, electron-phonon coupling (EPC) should be the dominant scattering mechanism for surface electronic states at finite temperatures. Hence, the study of EPC in TIs is of exceptional importance in assessing any potential applications. In this article both experimental and theoretical studies of the EPC on the surface of TIs are reviewed, with the contents mainly focused on the typical strong three dimensional TIs, such as $Bi_2Se_3$ and $Bi_2Te_3$.

Table of Contents

# 1. Introduction

## 1.1 Topological Insulators

Topological insulators (TIs) are a new class of materials in which a bulk gap for electronic excitations is generated because of the strong spin-orbit coupling that inverts bands of opposite parity close to the Fermi level. These materials are distinguished from ordinary insulators by the presence of gapless metallic surface states, resembling chiral edge modes in quantum Hall systems, but with unconventional spin textures. These exotic metallic states are formed by topological conditions that also render the electrons traveling on such surfaces insensitive to scattering by impurities.

### 1.1.1 Topological States

Before 1980, the different states of condensed matter, such as crystalline solids, magnets, and superconductors were classified by the symmetry they spontaneously break. In these examples we have broken translational, rotational and gauge symmetry, respectively.

The discovery of the Quantum Hall State in 1980 presented the first example of a quantum state that has no spontaneously broken symmetry. Its behavior depends on its topology and not on its specific geometry; it was topologically distinct from all previously known states of matter.

When electrons are subject to a large external magnetic field, electronic excitation gaps are generated in the sample bulk confining electrons to localized Landau orbitals. However, because of the confining potentials at the sample edges metallic conduction is permitted at the boundary, hence the birth of topologically robust chiral edge states [1]. The robustness arises because edge states are relegated to moving in a particular spatial direction at a given edge, with opposite edges carrying current in opposite directions. Hence an electron in an edge state cannot be backscattered by an impurity for there are no states of opposite momentum accessible to it.

Recently a distinct, albeit similar, topological phase of matter known as the Quantum Spin Hall phase was discovered [2, 3, 4, 5, 6]. One can think of this as two copies of the Quantum Hall State superimposed on each other. Any given edge supports electronic transport in both directions, but the edge states form degenerate Kramer's pairs such that the spins of oppositely propagating edge states are anti-aligned. In this case no external magnetic field is required and thus time reversal symmetry (TRS) is preserved.

One of the important discoveries of the past few years is that topological order also occurs in some three-dimensional materials. In these materials, the role of the magnetic field, which breaks TRS, is assumed by the mechanism of spin-orbit coupling, which maintains TRS, with spins of opposite sign counter-propagating along the surface boundaries. Thus, these materials actually have a closer analogy with the Quantum Spin Hall phase rather than the Quantum Hall effect. They have been named topological insulators [7, 8] because they are insulators in the bulk but have exotic metallic states present at their surfaces owing to the topological order.

We can simply describe a TI as one that always has a metallic boundary when placed next to a vacuum or an ordinary insulator. These metallic boundaries originate from topological invariants, which cannot change as long as a material remains insulating. In a normal insulator the Fermi level lies inside the energy gap separating the valence and conduction bands. However, the surface termination can introduce surface states in the energy gap. In most situations these conducting surface states are very fragile to any perturbation at the surface. In contrast, in a TI, these surface states are protected, that is, their existence does not depend on how the surface may be distorted. The reason for this is that the Hamiltonian describing a topological insulator and that describing the vacuum (which can be thought of as a type of insulator with an energy gap corresponding to particles and anti-particles) are topologically distinct. In particular one cannot transition between the two without closing the energy gap. It is this requirement that leads to the protected conducting surface states.

The distinction between an ordinary insulator and a TI lies in their respective sensitivity to boundary conditions (BCs). The former is strongly sensitive while the latter is insensitive to BCs. The two classes of insulators are characterized by their respective value of a nontrivial topological invariant identified as $Z_2$: $Z_2 = +1$ for ordinary insulators and $Z_2 = -1$ for TIs [3, 9, 10, 11, 12].

### 1.1.2 Dirac Fermion Character of the Surface State

The surface states in TIs fall in bulk energy gaps, and exhibit a single Dirac cone band dispersion [13, 14, 15, 16, 17], as shown in Fig.1.**a**. In two-dimensional ***k***-space, the dispersion relation looks like two cones that meet at a single point known as the Dirac point. The existence of an odd number of Dirac cones (here one) on the surface is ensured by the $Z_2$ topological invariant of the bulk. Kramers theorem dictates that for time-reversal invariant systems, the degeneracy of states with an even number of electrons (even number of cones) will always be lifted, but for odd electron number it is strictly maintained. For this reason, the surface states in TIs are said to be topologically protected.

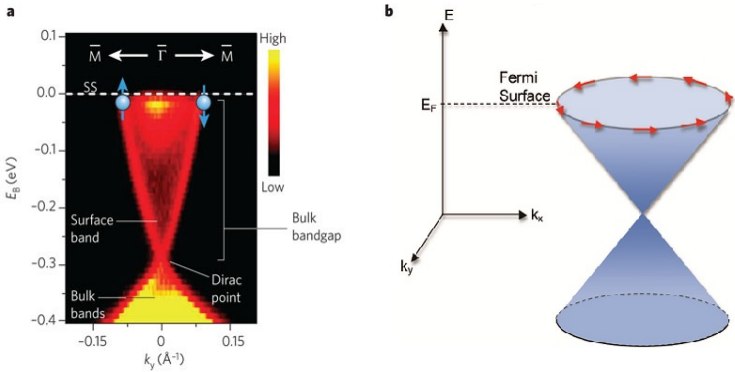

**Figure 1 a**, The electronic structure of a typical TI obtained from ARPES measurements. The direction of electron spin is indicated by blue arrows (the figure is adapted from Ref. [8], reprinted by permission from Macmillan Publishers Ltd: [Nature], copyright (2010)). **b**, Theoretical idealization of the electronic structure of a TI, showing the rotation of the spin degree of freedom (red arrows) as an electron moves around the Fermi surface or any constant.

Moreover, the strong spin-orbit coupling leads to a definite helicity whereby the spin is locked normal to the wave vector of the electronic state, as shown in Fig.1.**b**. A fundamental constraining feature of such spin-textured surface states is their robustness against spin-independent scattering, an attribute that protects them from backscattering and localization [18, 19]. This is because no time-reversal-invariant perturbation can open up an insulating gap at the Dirac point on the surface, owing to Kramers theorem.

The linear dispersion of the Dirac cone is reminiscent of that of light. In particular the electronic quasiparticles (collective excitations of many electrons) behave like photons but with a speed 300 times slower. People are interested in exploring the properties of massless electronic quasiparticles interacting with a lattice comprised of massive ions (phonons). This kind of electron-phonon interaction should have a distinguishable signature on both the electronic structure and the dynamical properties of the surfaces of these TIs.

### 1.1.3 Typical Strong Three Dimensional Topological Insulator Materials: $Bi_2Se_3$ and $Bi_2Te_3$

For a TI to form, spin-orbit coupling must be strong enough to modify the electronic structure significantly. Therefore, heavy-element, small-bandgap semiconductors are the most promising candidates. The search for TIs culminated in the recent discovery of TI behavior in $Bi_2Se_3$ and $Bi_2Te_3$ [13, 14].

$Bi_2Se_3$ has a non-trivial energy gap of 300 meV, larger than the energy scale of room temperature, while $Bi_2Te_3$ has an indirect gap of about 150 meV. The surface states for these crystals are extremely simple, described by a single gapless Dirac cone centered at the $k = 0$, $\bar{\Gamma}$ point in the surface Brillouin zone (SBZ). These crystals share the same rhombohedral structure which belongs to space group $D_{3d}^5$ ($R\bar{3}m$) with five atoms per unit cell. The crystal has a trigonal axis in the [111]-direction (three-fold rotation symmetry), defined as the z axis, a two-fold rotation axis, defined as the x axis, and a diagonal axis (oriented 30 degrees from the reflection plane), defined as the y axis.

The crystal structure of $Bi_2Se_3$ is shown as an example in Fig.2. It has a layered structure with a triangular lattice within each layer. Sets of five atomic planes stack along the z-direction to form the so-called quintuple layers (QLs). Each QL consists of five atoms with two equivalent Se atoms (Se1), two equivalent Bi atoms and a third Se atom (Se2), ordered in the Se1-Bi-Se2-Bi-Se1 sequence . The Se2 site is an inversion center, through which Bi and Se1 atoms are interchanged, respectively. The coupling is strong between two atomic layers within one QL but much weaker, predominantly of the van der Waals type, between successive QLs. The primitive lattice vectors in the rhombohedral and hexagonal basis are:

$$t_1^R = (-\frac{a}{2}, -\frac{\sqrt{3}a}{6}, \frac{c}{3}), \ t_2^R = (\frac{a}{2}, -\frac{\sqrt{3}a}{6}, \frac{c}{3}), \ t_3^R = (0, \frac{\sqrt{3}a}{6}, \frac{c}{3}),$$

$$t_1^H = (\frac{\sqrt{3}a}{2}, -\frac{a}{2}, 0), \ t_2^H = (0, a, 0), \ t_3^H = (0, 0, c),$$

where *a* and *c* are lattice constants of the hexagonal cell. The corresponding reciprocal vectors are given as:

$$\boldsymbol{b}_1^R = (-1, \frac{\sqrt{3}}{6}, \frac{a}{c})\frac{2\pi}{a}, \quad \boldsymbol{b}_2^R = (1, -\frac{\sqrt{3}}{6}, \frac{a}{c})\frac{2\pi}{a}, \quad \boldsymbol{b}_3^R = (0, \frac{2\sqrt{3}}{6}, \frac{a}{c})\frac{2\pi}{a}$$

$$\boldsymbol{b}_1^H = (\frac{2}{\sqrt{3}}, 0, 0)\frac{2\pi}{a}, \quad \boldsymbol{b}_2^H = (\frac{1}{\sqrt{3}}, 1, 0)\frac{2\pi}{a}, \quad \boldsymbol{b}_3^H = (0, 0, 1)\frac{2\pi}{c}.$$

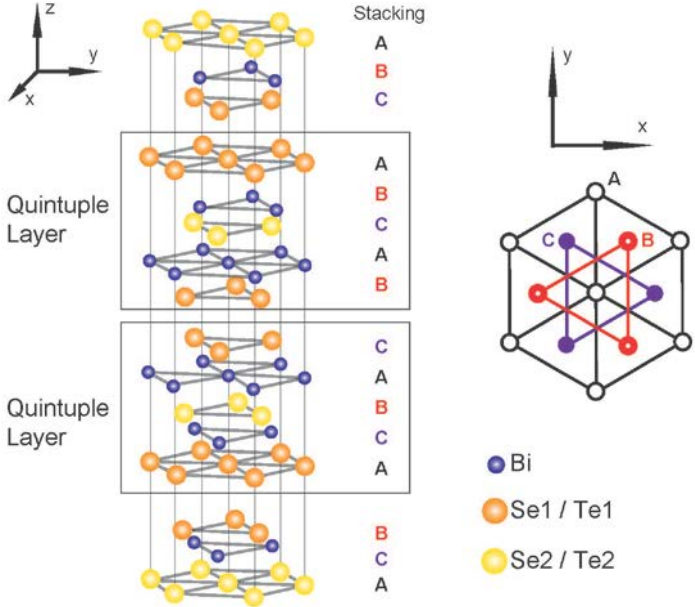

**Figure 2** Crystal Structure of $Bi_2Se_3$ and $Bi_2Te_3$. Left side is a unit cell showing the quintuple structure. Right side is a top view showing the stacking arrangement.

The bulk Brillouin zone and surface Brillouin zone are shown in Fig.3.**a**, while an extended version of the SBZ is displayed in Fig.3.**b**.

Interested readers can find more detailed information about TIs in several recent reviews [20, 21, 22].

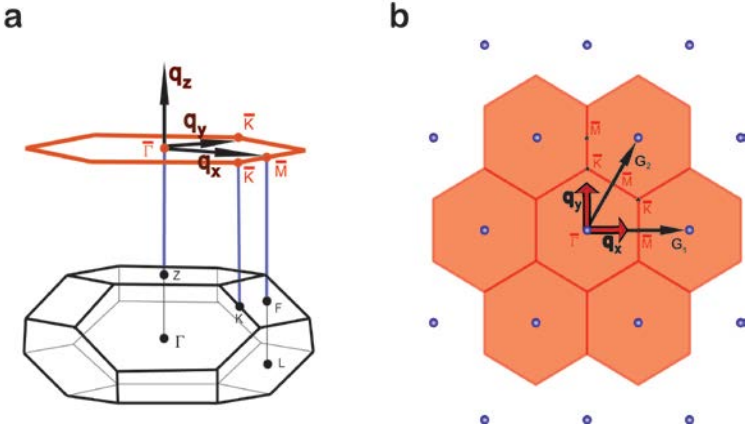

**Figure 3 a**, A schematic view of the bulk three-dimensional BZ of Bi$_2$Se$_3$ (Bi$_2$Te$_3$) and the two-dimensional BZ of the projected (001) surface. **b**, The extended surface BZ with reciprocal lattice vectors **G$_1$** and **G$_2$**. the two high symmetry directions are labeled as $\overline{\Gamma}-\overline{M}$ and $\overline{\Gamma}-\overline{K}$, which are aligned along **q$_x$** and **q$_y$**, respectively.

## 1.2 Electron-Phonon Coupling

The interactions between electrons and the collective excitations of bosons are dominant driving forces in some remarkable properties emerging from many-body effects in materials. As one of the most important interactions, electron-phonon coupling (EPC) plays a vital role in many physical properties ranging from heat capacity and electric conductivity to superconductivity. EPC changes the dispersion and the lifetime of both the electronic and phonon states in a material. In theory, EPC induced renormalization of the electronic structure and phonon bands can be reflected by the complex electron self-energy $\Sigma$ and phonon self-energy $\Pi$. The real part Re$\Sigma$ (or Re$\Pi$) renormalizes the electron (phonon) energy dispersion, while the imaginary part Im$\Sigma$ (or Im$\Pi$) accounts for the finite lifetime of the electron (phonon) state arising from the interaction. Usually it is the electron's effective mass and dispersion close to the Fermi surface that is modified by the EPC, as illustrated in Fig.4. The increase of the effective mass is characterized by the EPC mass enhancement factor $\lambda$ with m* = $m_0(1+\lambda)$, where m* and $m_0$ are the effective masses with and without EPC, respectively. $\lambda$ is commonly used to describe the strength of EPC, which can be extracted by measurements of electrical transport and thermal transport [23].

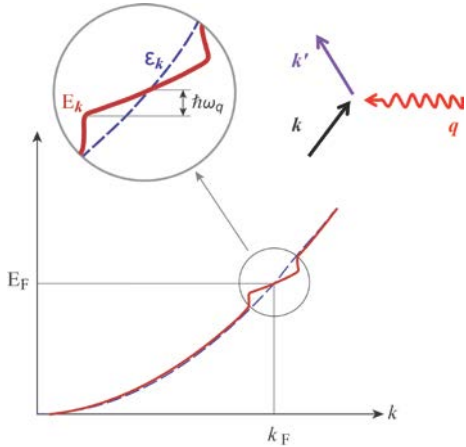

**Figure 4** Renormalization of the electronic dispersion due to EPC (schematic). The blue dashed line is the bare dispersion $\varepsilon(k)$; the red solid line is the renormalized dispersion E(k). Inset left: magnification picture close to the Fermi energy. Inset right: illustration of the electron's transition from initial state ***k*** to final state ***k′*** via a phonon with energy $\hbar\omega_q$ and wave vector ***q***.

$\lambda$ can also be expressed by a more basic characteristic function called the Eliashberg function (ELF), $\alpha^2F(\omega)$, which is a product of an effective EPC strength $\alpha^2$ involving phonons with energy $\hbar\omega$ and the phonon density of states $F(\omega)$. It determines the mass enhancement factor with the expression of

$$\lambda = 2\int_0^{\omega_{max}} \frac{\alpha^2 F(\omega)}{\omega} d\omega. \qquad (1)$$

With $\alpha^2 F(\omega)$, one can distinguish the contributions of phonon with different energy $\omega$, not the average of all phonon modes. Several experimental techniques such as tunneling spectroscopy or heat capacity measurements can determine the function $\alpha^2 F(\omega)$ [23].

The details of how to determine $\lambda$ and $\alpha^2 F(\omega)$ on the surfaces of metals have been reviewed elsewhere [24, 25, 26].

## 2. Electron-phonon Coupling on the Surface of Topological Insulators

Surface Dirac fermions in TIs are robust against backscattering, but other scattering events can affect their anticipated ballistic behavior. Technical improvements may minimize or eventually eliminate surface defects, but phonons are always present. Consequently, coupling to phonons should be the dominant scattering mechanism for Dirac fermions on these surfaces at finite temperatures. Indeed, considerable efforts have been made recently by the condensed matter community to study the EPC on the surface of TIs. Experimental studies have been carried out by angle resolved photoemission spectroscopy (ARPES) and helium atom scattering (HAS). The former examines the renormalization of the surface electron energy dispersions in the vicinity of Fermi surface caused by e-p interactions, and hence deduces the strength of the EPC. By contrast, HAS studies the renormalization of the surface phonon dispersions due to e-p interactions, and determines mode-specific surface EPC. Theoretically, investigations of the EPC mechanism were based on both analytical models and *ab initio* calculations. However, a consensus about the EPC strength on the surface of TIs has not been achieved, since conflicting values have been reported. In this section, both experimental and theoretical studies of the EPC on the surface of TIs are reviewed, with the contents mainly focused on the typical strong three dimensional TIs, such as $Bi_2Se_3$ and $Bi_2Te_3$.

## 2.1 Experimental Studies by ARPES

### 2.1.1 Brief Introduction to ARPES

ARPES, a unique experimental tool based on the photoelectric effect, provides direct access to electronic band structure and many-body effects in solids. The many-body effects of the photoemission process are manifest in the finite life time, τ, of the electronic state due to possible scattering mechanisms such as electron–electron (ee) scattering, electron–phonon (ep) scattering or the interaction with lattice impurities (im). The life time is related to the photoelectron spectroscopy line width, Γ, by the relation

$$\Gamma = \frac{\hbar}{\tau} = \Gamma_{ee} + \Gamma_{ep} + \Gamma_{im}. \qquad (2)$$

It has been shown that the photoelectron intensity is proportional to the so-called spectral function

$$A(\boldsymbol{k}, E) = \frac{1}{\pi} \frac{\left|\mathrm{Im}\,\Sigma(\boldsymbol{k}, E)\right|}{[E - \varepsilon(\boldsymbol{k}) - \mathrm{Re}\,\Sigma(\boldsymbol{k}, E)]^2 + [\mathrm{Im}\,\Sigma(\boldsymbol{k}, E)]^2} f(E), \qquad (3)$$

where $\boldsymbol{k}$ is the electron wave vector, $\varepsilon(\boldsymbol{k})$ is the bare electron dispersion if no interaction is present, $\Sigma(\boldsymbol{k}, E)$ is the complex electron self-energy, and *f(E)* denotes the Fermi-Dirac distribution function. In spectroscopic terms, the relation $\Gamma = -2\,\mathrm{Im}\,\Sigma(\boldsymbol{k}, E)$ is frequently used. Because the real and imaginary parts of the self-energy are related by a Hilbert (or Kramers-Kronig) transformation, it is sufficient to determine either $\mathrm{Re}\,\Sigma(\boldsymbol{k}, E)$ or $\mathrm{Im}\,\Sigma(\boldsymbol{k}, E)$.

A typical ARPES experiment facility consists of a monochromatic light source, e.g. a gas discharge lamp, an x-ray tube or a synchrotron radiation source, and a spherical-deflection type electrostatic analyzer. In a general experimental setup, the directions of the light source and the analyzer are usually fixed, while the sample can be rotated in both the azimuthal and polar directions. For a fixed photon energy and a fixed emission angle, the obtained photoemission intensity *I (E)* is a function of kinetic energy, which is called the energy distribution curve (EDC). The data set obtained at constant electron kinetic energy is a function of momentum, which is called the momentum distribution curve (MDC). The methods that have been applied to extract the self-energy from photoemission data are closely related to those two measuring modes of ARPES. Line positions and shapes of an electronic band with weak dispersion (flat band) are best analyzed using EDCs since the angle resolution of the spectrometer can be neglected. In contrast, the analysis of line positions and shapes of electronic bands with strong dispersion (steep bands) is facilitated using MDCs in order to minimize the effect of angular resolution of the instrument [25, 26]. The details of how to obtain the electron self-energy from ARPES measurements can be found elsewhere [27, 28].

To extract the pure EPC contribution $\Gamma_{ep}$ to the measured total line width $\Gamma$, one should notice the fact that the temperature dependence of both $\Gamma_{ee}$ and $\Gamma_{im}$ is usually very small. However, $\Gamma_{ep}$ is strongly temperature-dependent. By measuring the electron self-energy in the vicinity of the Fermi energy ($E_F$) as a function of temperature, one can fit the ensuing temperature-dependence of lifetime broadening $\Gamma_{ep}$ to the relation [23]

$$\Gamma_{ep}(\boldsymbol{k}, E) = 2\pi\lambda(\boldsymbol{k}, E)k_B T \qquad (4)$$

valid at high temperatures (when $k_B T$ is higher than the maximum phonon energy), and extract the value of EPC parameter $\lambda$.

Another technique commonly used to determine the parameter $\lambda$ is to calculate the slope of the real part of the electron self-energy Re$\Sigma$ at the Femi energy [24]

$$\lambda = -\frac{d\,\text{Re}\,\Sigma(\varepsilon)}{d\varepsilon}\bigg|_{E_F}. \qquad (5)$$

But Re$\Sigma(\varepsilon,T)$ is usually temperature-dependent and determining the slope of the bands or Re$\Sigma(E_F)$ at the Fermi energy is difficult due to the fact that the photoemission spectra are cut off at the Fermi energy [24].

### 2.1.2 Experimental Results from ARPES Measurement

As the primary technique for obtaining the surface electronic structure of solids, ARPES has been used to characterize the surface topological states of TIs since their initial discovery [14, 15, 16, 17, 18]. More recently people have begun to extract the electron self-energy from ARPES data to determine the strength of EPC on the surface of TIs.

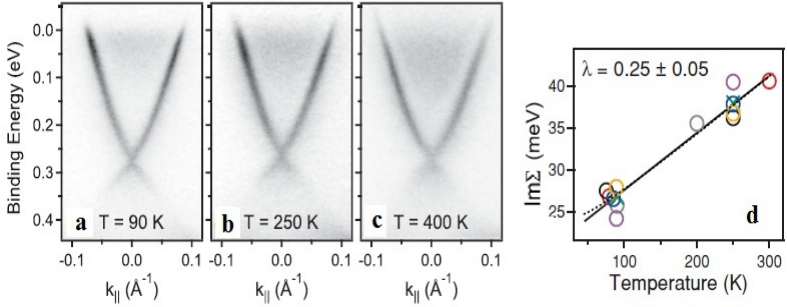

**Figure 5** ARPES measurements of $Bi_2Se_3$ from Hofmann's group. **a**, **b** and **c**, ARPES spectra taken at T = 90 K, T = 250 K, and T = 400 K, respectively. **d**, Imaginary part of the extracted electron self-energy ImΣ as a function of temperature. The solid line is a linear fit to the experimental data and the dotted line shows a calculation for a full Debye model. (Figures are reprinted with permission from Ref. [29]. Copyright (2011) by the American Physical Society.)

Fig.5 shows the results from Hofmann's group using synchrotron radiation ARPES [29]. They concentrated on the spectral region in which only the topological state exists, and, thus, only intraband scattering is possible. Photoemission spectra were taken between 90 and 400 K, and the imaginary part of the electron self-energy ImΣ was extracted from the spectra. Then ImΣ was plotted as a function of temperature. By making use of Equation (4), a linear fit of the data provided the value of the EPC parameter to be $\lambda = 0.25 \pm 0.05$. They claimed that this unexpected large value implies a very significant EPC for the topological state.

Fig.6 shows the results of another recent study from Valla's group using synchrotron radiation ARPES as well [30]. They also performed temperature-dependent measurements of the electronic dispersions and extracted the imaginary part of the electron self-energy ImΣ. Again, a linear fit of the experimental data yielded the value of EPC parameter, $\lambda \cong 0.08$, which they claimed to be one of the weakest coupling constants ever reported in any material. They also used a magnified image (Fig.6.**c**) of the low energy region of the ARPES spectrum, as supporting evidence, to show the absence of a sudden change in the slope or a "kink" in the electronic dispersion. This result was consistent with an earlier ARPES study in Ref.[31], where the EPC was found to be very weak but the coupling constant λ was not calculated.

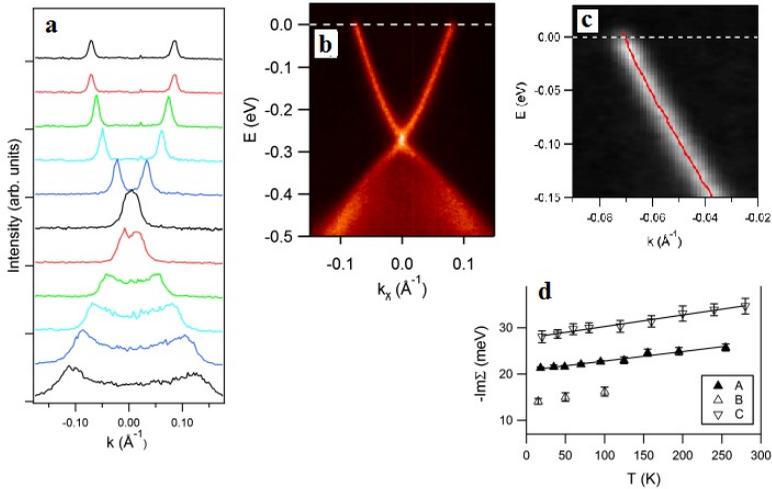

**Figure 6** ARPES measurements of $Bi_2Se_3$ from Valla's group. **a**, MDCs corresponding to the spectrum shown in panel (**b**), spaced by 50 meV, with the top curve representing the Fermi level. **b**, ARPES spectrum along the $\overline{\Gamma} - \overline{K}$ line in the SBZ taken at 18 K. **c,** Zoom in of the low-energy region of the ARPES spectrum panel (**b**). **d**, Imaginary part of the extracted electron self-energy ImΣ as a function of temperature for three different samples. (Figures are reprinted with permission from Ref. [30]. Copyright (2012) by the American Physical Society.)

An obvious disadvantage of the above two studies using ARPES based on a synchrotron radiation source is that they can only attain a resolution of about 10 meV. This energy resolution limits the detection energy range, which makes it impractical to explore the possible low energy interactions smaller than 10 meV.

With the development of instrumentation, ultraviolet lasers can now be used as the photon source for ARPES. The energy resolution of the first laser-based ARPES is better than 1 meV [32]. A more recent ARPES apparatus can even achieve a maximum energy resolution of 70 μeV and cooling temperature of 1.5 K [33]. Using the high resolution laser-based ARPES, it is possible to explore the details of the EPC on the surface of TIs in the low energy (several meV) range.

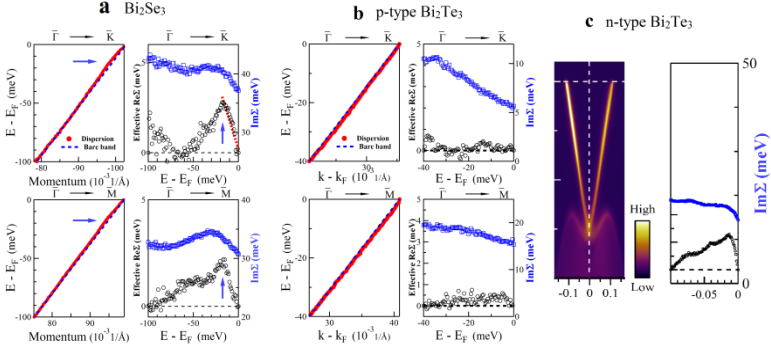

**Figure 7** ARPES measurements of $Bi_2Se_3$, p-type $Bi_2Te_3$ and n-type $Bi_2Te_3$ from Zhou's group. **a,** Electronic dispersion and the extracted electron self-energy along both $\overline{\Gamma} - \overline{K}$ and $\overline{\Gamma} - \overline{M}$ directions for $Bi_2Se_3$. **b,** Electronic dispersion and the extracted electron self-energy along both $\overline{\Gamma} - \overline{K}$ and $\overline{\Gamma} - \overline{M}$ directions for p-type $Bi_2Te_3$. **c,** Electronic dispersion and the extracted electron self-energy along both $\overline{\Gamma} - \overline{K}$ direction for n-type $Bi_2Te_3$. (Figures are adapted from Ref. [34])

Fig.7 shows the results from Zhou's group using laser-based ARPES. They measured the electronic dispersion of the TIs $Bi_2(Te_{3-x}Se_x)$, with varying x from 0 to 3, where $Bi_2Se_3$, p-type $Bi_2Te_3$, and n-type $Bi_2Te_3$ are all included. Both the real part and the imaginary part of the electron self-energy are extracted from the spectra. The EPC constants are determined by calculating the slope of the real part ReΣ near the Fermi energy using Equation (5). They obtained the following results: for $Bi_2Se_3$, $\lambda \cong 0.17$; for p-type $Bi_2Te_3$, $\lambda \cong 0$; and for n-type $Bi_2Te_3$, $\lambda \cong 0.19$ [34].

Fig.8 shows the results from Shin's group using laser-based ARPES as well. They obtained the photoemission spectra at 7 K for $Bi_2Se_3$, $Bi_2Te_3$ and Cu-doped $Bi_2Se_3$. They found two "kinks" in the real part of the extracted electron self-energy at ~ -15 and ~ -3 meV, which indicate

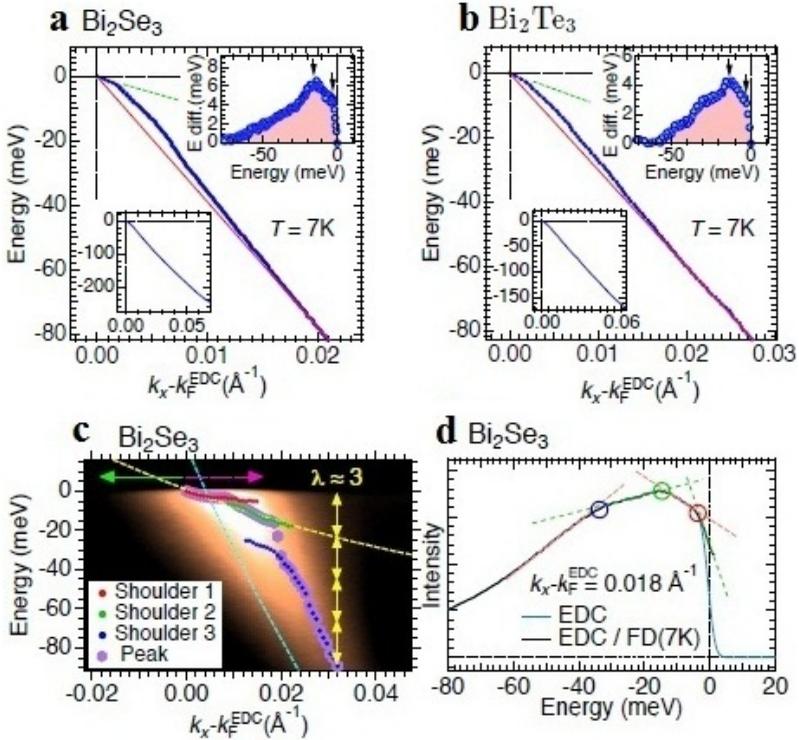

**Figure 8** ARPES measurements of $Bi_2Se_3$ and $Bi_2Te_3$ from Shin's group. **a**, Electronic dispersion and the extracted electron self-energy along $\overline{\Gamma} - \overline{K}$ direction for $Bi_2Se_3$. **b**, Electronic dispersion and the extracted electron self-energy along $\overline{\Gamma} - \overline{K}$ direction for $Bi_2Te_3$. **c,** Parabolic bands (dashed lines) with a mass of $0.83m_e$ and $0.14m_e$ superimposed on the APRES image of $Bi_2Se_3$. **d**, Typical EDC of panel (**c**) with three electron-boson interaction features marked. (Figures are adapted from Ref. [37])

electron couplings with two different kinds of collective modes. The 15 meV anomaly was attributed to the coupling with high energy $A_{1g}$ optical phonon modes, whereas the 3 meV one was attributed to the coupling with both the low energy optical surface phonon [35] and the theoretically proposed "spin-plasmon" [36]. By calculating the slope of the real part $Re\Sigma$ near the Fermi energy using Equation (5), they obtained an exceptionally large coupling constant of $\lambda \cong 3$, which they claimed to be the strongest ever reported for any material [37].

## 2.2 Experimental Studies by HAS

### 2.2.1 Brief Introduction to HAS

HAS, the counterpart of neutron scattering on surfaces, has been successfully used to explore surface structures and surface lattice dynamics since the early 1980s [38, 39, 40, 41]. As a surface probe, He atomic beams have a number of advantages over other conventional surface probes such as electrons. The key features of HAS are the following [42]:

- Thermal He atoms are inert and non-destructive probes, because they are neutral low energies particles.
- Thermal HAS is an exclusively surface sensitive technique, since the classical turning point usually lies at about 2-3 Å from the surface plane, defined by the surface ion core centers.
- The characteristic de Broglie wavelengths of the beam are in the range 0.5 to 1.5 Å, which is comparable to the lattice constants in solids.
- The He beams employed have very high energy resolution (< 1 meV), and very high intensity, of the order of $10^5$ -$10^6$ atoms/s.
- The thermal energy of the He beam is between 10 and 100 meV and the momentum exchange can cover the entire SBZ. These energies are well matched to that of the surface phonons.
- Because the He atom is very light compared to other inert gas atoms, multi-phonon events are generally excluded in HAS measurements.

A typical HAS experiment facility consists of a high-pressure nozzle beam production system, which generates a monoenergetic He beam, and a detector that can pulse the He beam and apply time-of-flight techniques to determine the beam energy. Measurements of diffraction spectra (elastic scattering) allow not only the determination of the size and orientation of the surface unit cells but also, by means of analyzing diffraction intensities, yield the surface corrugation which enables one to image direct pictures of the geometrical arrangement of the surface atoms [38]. The surface lattice dynamics, or the dispersion of the surface phonons, are determined by analyzing inelastic scattering events.

Electron-Phonon Coupling on the Surface of Topological Insulators    19

In the inelastic case, the energy and momentum of the surface phonons are involved. Energy conservation of the entire system, consisting of the incoming He atoms and the whole crystal, requires that

$$\Delta E = \hbar\omega(\mathbf{Q}) = E_f - E_i = \frac{\hbar^2}{2m}(k_f^2 - k_i^2), \quad (6)$$

where $\hbar\omega(\mathbf{Q})$ is the energy of the phonon with wave vector $\mathbf{Q}$ projected onto the surface plane, $E$ denotes the beam energy and $k$ denotes its wave vector. The subscripts $i$ and $f$ represent the incoming (incident) and outgoing (scattered) waves respectively.

Conservation of momentum for in-plane scattering can be written as:

$$\Delta K = |\mathbf{G} + \mathbf{Q}| = k_f \sin\theta_f - k_i \sin\theta_i, \quad (7)$$

where $\mathbf{G}$ is a surface reciprocal lattice vector, and $\theta_i$ ($\theta_f$) represents the angle between the incident (outgoing) beam and the surface normal.

Once the energy difference $\Delta E$ between the incoming and outgoing beam is measured, the phonon dispersion can be determined using Equation (6) and (7). The details can be found in Ref. [39, 42].

It is well established that He atoms are scattered by the oscillations of the surface electron density about 2–3 Å away from the first atomic layer, and thus HAS is very sensitive to phonon-induced surface charge density oscillations, even those induced by subsurface second-layer displacements [43, 44]. Consequently, HAS intensities carry direct information on the surface charge density oscillations associated with surface phonons and ultimately on the surface electron-phonon interaction.

### 2.2.2 Experimental Results from HAS Measurement

Recent measurements of phonon dispersion curves on the (001) surfaces of several binary and ternary TIs were carried out using coherent inelastic HAS techniques. The results for $Bi_2Se_3$ are shown in Fig. 9. In order to interpret and fit the experimental inelastic data, phenomenological surface lattice dynamical calculations, based on the pseudocharge model (PCM) [45, 46] and applied to slab geometries containing 30 QLs were employed. The dispersion curves reveal several features: First, the absence of long-wavelength Rayleigh waves. Second, the appearance of a low-lying optical phonon branch with isotropic convex dispersive character in the vicinity of the $\overline{\Gamma}$-point. Lattice dynamics calculations based on the PCM show that the optical phonon branch appears with a concave shape when Dirac fermions are absent, but its dispersion changes to a convex shape when Dirac fermions are present. Moreover, this optical branch displays a V-shaped minimum at approximately $2k_F$ that defines a Kohn anomaly. The physical role of the surface Dirac fermion states in the phonon energy renormalization was subsequently studied with the aid of a microscopic model based on Coulomb-type perturbations within the random phase approximation (RPA). The model was incorporated in calculations of density-density correlations that take into account the helicity and linear dispersion of the surface Dirac fermions states. This dispersive profile was attributed to the renormalization of the surface phonon excitations by the surface Dirac fermions [35].

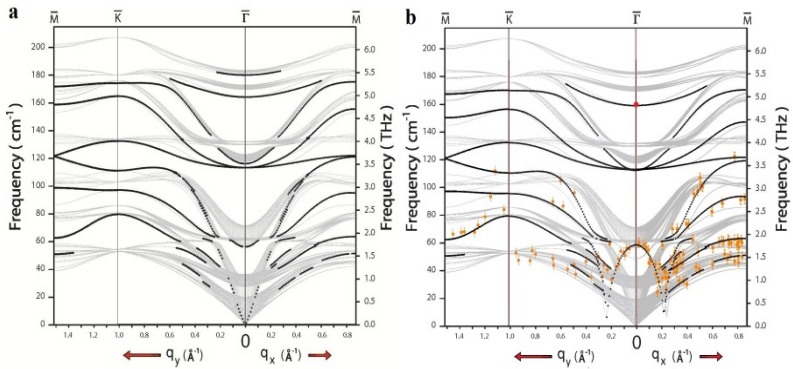

**Figure 9** HAS results of Bi$_2$Se$_3$ from El-Batanouny's group. **a**, Calculated surface phonon dispersion curves (black dots) along the $\overline{\Gamma} - \overline{M}$ and $\overline{\Gamma} - \overline{K}$ directions when the role of surface Dirac fermions is not accounted for. The gray area represents the projection of the bulk bands onto the SBZ. **b**, Surface phonon dispersion curves along the $\overline{\Gamma} - \overline{M}$ and $\overline{\Gamma} - \overline{K}$ directions when the role of surface Dirac fermions is accounted for: Experimental data appear as orange dots with error bars reflecting instrument resolution, while the calculated surface dispersion curves are represented by black dots. (Figures are reprinted with permission from Ref. [35]. Copyright (2011) by the American Physical Society.)

In order to obtain the strength of the EPC from the phonon measurements, EPC-induced phonon linewidths are needed. However, Phonon line broadenings due to EPC are generally small, even for superconductors with strong coupling. Such small broadenings are very difficult to detect in both neutron and helium scattering experiments mainly because of unavoidable instrument linewidths of a few meVs. This limits possibilities of extracting EPC contributions from the measured phonon linewidth. Moreover, in addition to EPC, there are other inherent contributions to phonon line broadening, such as phonon-phonon interaction (anharmonicity), phonon-defect scattering, and phonon anticrossing with other branches. Thus an indirect method was used to obtain the EPC. First the real part of the phonon self-energy was fit to the measured surface-phonon dispersion curve. The imaginary part, and hence the e-ph contribution to the phonon linewidth, was obtained with the aid of a Hilbert (or Kramers-Kronig) transform [47].

This approach was supported by the fact that the experimentally measured temperature independence of the dispersion demonstrates that the surface-phonon frequency renormalization is mainly determined by EPC. Panels (**a**), (**b**) and (**c**) of Fig.10 shows the measurements of the phonon dispersion as a function of temperature. It is obvious that the phonon dispersion is almost independent of temperature.

Renormalization of phonon frequencies due to the EPC is described by the Dyson equation

$$(\hbar\omega_q)^2 = (\hbar\omega_q^{(0)})^2 + 2\hbar\omega_q^{(0)} \, \text{Re}[\Pi(q,\omega_q)], \qquad (8)$$

where $\omega_q$ and $\omega_q^{(0)}$ are the renormalized and bare surface-phonon frequency, respectively, and $\Pi(q,\omega_q)$ is the corresponding phonon self-energy. $\text{Re}[\Pi(q,\omega_q)]$ was calculated using the best-fit parameters obtained in Ref. [35] at different wave vectors along the $\bar{\Gamma}-\bar{M}$ direction. The corresponding imaginary part of the phonon self-energy was obtained using the Hilbert transform

$$\text{Im}[\Pi(q,\omega_q)] = \frac{2}{\pi} \int_0^\infty \frac{\omega_q}{\omega_q^2 - \omega_q'^2} \text{Re}[\Pi(q,\omega_q')] d\omega_q'. \qquad (9)$$

The results for $\text{Re}[\Pi(\boldsymbol{q},\omega_q)]$ and $\text{Im}[\Pi(\boldsymbol{q},\omega_q)]$ are plotted in panels (**d**) and (**e**) of Fig.10 respectively. The phonon linewidth defined by $\gamma = -2\,\text{Im}\,\Pi$ is plotted in Fig.10.**f**.

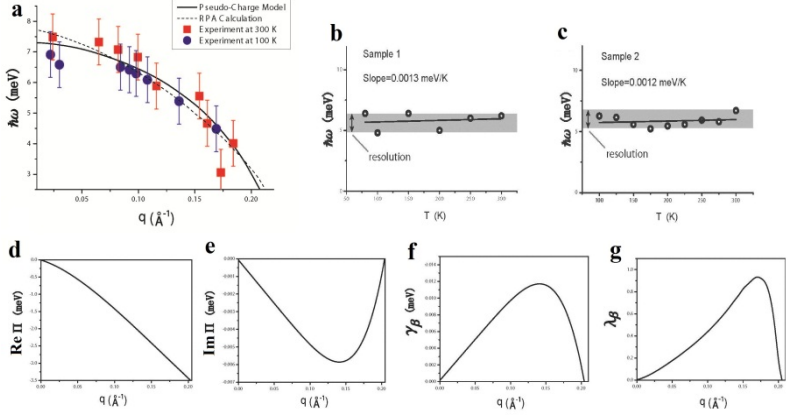

**Figure 10** HAS results of $Bi_2Se_3$ from El-Batanouny's group. **a**, Dispersion curve of the isotropic optical surface phonon branch. **b** and **c**, The energy of a single phonon event as a function of temperature for Sample 1 at $q \sim 0.13$ Å$^{-1}$ (**b**) and for Sample 2 at $q \sim 0.6$ Å$^{-1}$ (**c**), respectively. The circles represent the experimental data, while the solid line stands for a linear fit of the data. The gray band represents the energy resolution of the facility. **d**, The real part of the phonon self-energy $\text{Re}\Pi$ for the isotropic optical surface phonon branch. **e**, The imaginary part of the phonon self-energy $\text{Im}\Pi$. **f**, The EPC-induced linewidth $\gamma = -2\,\text{Im}\,\Pi$. **g**, the coupling parameter determined from $\gamma$. (Figures are reprinted with permission from Ref. [47]. Copyright (2012) by the American Physical Society.)

The corresponding EPC constant was obtained with the aid of the expression

$$\lambda(\boldsymbol{q}) = \frac{1}{2\pi N(E_F)} \frac{\gamma(\boldsymbol{q})}{(\hbar\omega_q)^2}, \qquad (10)$$

where $N(E_F)$ is the electronic density of states at $E_F$. The results are plotted in Fig.10.**g**. Averaging over the function $\lambda(q)$, an effective EPC constant value for the dispersive optical branch was obtained $\lambda \cong 0.43$ [47].

## 2.3 Theoretical Studies

### 2.3.1 Analytical Phonon Models

The surface acoustic phonon propagation in TIs has been investigated using the continuum limit for the surface conduction state and the phonon modes [48]. This study assumed a linear dispersion in both cases given by Fermi velocity $v_F$ and sound velocity $v_s$ respectively. The low energy effective Hamiltonian was written as

$$H_0 = iv_F \int d^2\mathbf{r}\psi^{\dagger}(\mathbf{r})\boldsymbol{\sigma} \cdot (\hat{\mathbf{z}} \times \nabla)\psi(\mathbf{r}), \quad (11)$$

where $\boldsymbol{\sigma}$, and $\hat{\mathbf{z}}$ are the electron spin and the fixed surface normal, respectively, and $\psi(\mathbf{r})$ is a two component spinor for the Dirac electrons given by the expression

$$\psi(\mathbf{r}) = \frac{1}{\sqrt{V}}\int d^2\mathbf{r} e^{-i\mathbf{k}\cdot\mathbf{r}}\psi_\mathbf{k}; \quad \psi_\mathbf{k} = \begin{pmatrix} c_{\mathbf{k}\uparrow} \\ c_{\mathbf{k}\downarrow} \end{pmatrix}. \quad (12)$$

An adiabatic generalization of the unperturbed Hamiltonian in Equation.(12), valid for $v_s \ll v_F$, was done by replacing the global surface normal $\hat{\mathbf{z}}$ by the local normal $\hat{n}(\mathbf{r})$. This leads to a modified low-energy and long-wavelength effective Hamiltonian for the coupling of surface phonons and helical Dirac states:

$$H_0 = iv_F \int d^2\mathbf{r}\psi^{\dagger}(\mathbf{r})\boldsymbol{\sigma} \cdot [\hat{n}(\mathbf{r}) \times \nabla]\psi(\mathbf{r}). \quad (13)$$

For small phonon wave vectors and amplitudes, $\hat{n}(\mathbf{r})$ can be replaced by a first order expansion $\hat{n}(\hat{\mathbf{r}}) = \mathbf{z} + \delta n(\mathbf{r})$, where $\delta \hat{n}(\mathbf{r})$ is the local deviation from the global surface normal due to the phonon mode. The geometry of this deformation is schematically shown in Fig.11.**a**.

Using this model, an effective EPC term of the Hamiltonian was derived and the phonon dispersion curve was calculated. The resulting renormalized phonon frequency change due to EPC is plotted in Fig.11.**b**. Cleary the surface phonon dispersion exhibits a Kohn anomaly near $2k_F$. This result seems consistent with the experimental HAS measurements mentioned above. However, the HAS experimental data demonstrate the absence of acoustic surface-phonon modes, and the discovered Kohn anomaly is in an optical phonon branch [35].

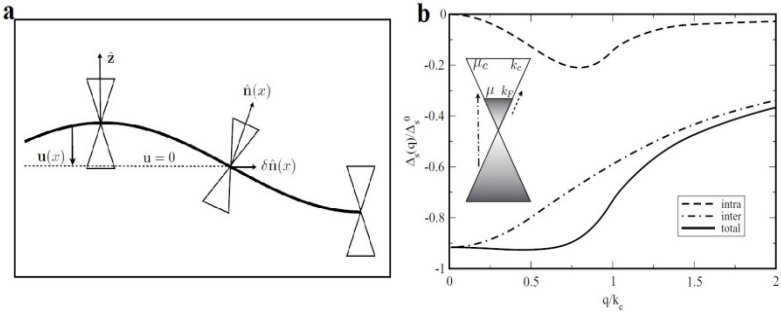

**Figure 11** Theoretical model from Thalmeier. **a**, Schematic view of adiabatic modification of helical surface states by a surface phonon. The surface normal $\hat{n}(\mathbf{r}) = \hat{\mathbf{z}} + \delta \hat{n}(\mathbf{r})$ giving the local orientation of the Dirac cone is periodically modulated by the surface displacement amplitude **u**(x). **b**, Renormalized phonon frequency change as a function of phonon wave vector, with $k_c=2k_F$. Interband contribution is monotonic while intraband part exhibits a Kohn anomaly for $q/k_c = q/2k_F < 1$. The inset shows a schematic view of intra- and interband excitation processes. (Figures are reprinted with permission from Ref. [48]. Copyright (2011) by the American Physical Society.)

Another similar e-p scattering theory for TIs was recently proposed based on an isotropic elastic continuum phonon model with stress-free boundary conditions [49]. In this theory, phonons are modeled to interact with the

Dirac surface fermions via the deformation potential. The Hamiltonian of the system is

$$H = H_e + H_p + H_{ep}, \qquad (14)$$

where $H_e$ is the massless 2D Dirac electron term, $H_p$ is the noninteracting phonon term, and $H_{ep}$ is the EPC term. For simplicity, I will not list the explicit form of these terms here in this review. They are formulated in Ref.[49]. Using this model, the EPC induced lifetime broadening and the Eliashberg function for the acoustic phonon modes were calculated. The EPC constants were obtained using Equation.(1). The calculated results are shown in Fig.12.

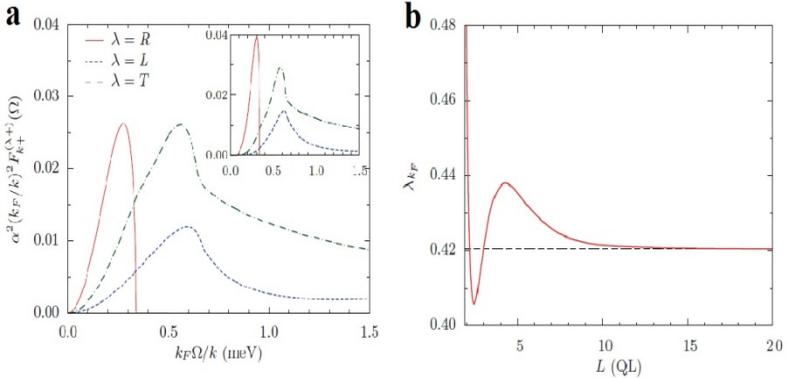

**Figure 12** Theoretical model from Giraud and Egger. **a**, Calculated low-frequency behavior of the Eliashberg functions for the three relevant acoustic phonon modes R, L and T. The results are shown for a semi-infinite geometry (Figure is reprinted with permission from Ref. [49]. Copyright (2011) by the American Physical Society.). **b**, Width ($L$) dependence of the effective EPC constant at the Fermi level for thin films. The dashed horizontal line indicates one-half of the effective coupling constant in the semi-infinite geometry. (Figure is reprinted with permissions from Ref. [50]. Copyright (2012) by the American Physical Society.)

This study yielded several EPC constants:

1. $\lambda$=0.13 for $Bi_2Te_3$ with a semi-infinite geometry [49].
2. $\lambda$=0.84 for $Bi_2Se_3$ with a semi-infinite geometry [50].
3. $\lambda$=0.42 for $Bi_2Se_3$ thin films [50].

### 2.3.2 *ab initio* Calculations

Lattice dynamical calculations using *ab initio* quantum-mechanical techniques based on density functional perturbation theory are the state of the art of computational materials science. It enables one to calculate lattice vibrations of materials with only the information of the chemical compositions [51].

Recently, surface-phonon dispersions and e-p interactions in $Bi_2Te_3$ films were calculated using density functional perturbation theory with spin-orbit coupling included [52]. The phonon density of states and the localized surface phonon modes along $\overline{\Gamma}-\overline{M}$ and $\overline{\Gamma}-\overline{K}$ directions in the SBZ were obtained. The results are shown in Fig.12. The ELF $\alpha^2 F(\omega)$ was also calculated. Using Equation.(1), the EPC constant for $Bi_2Te_3$ films yielded λ ~ 0.05.

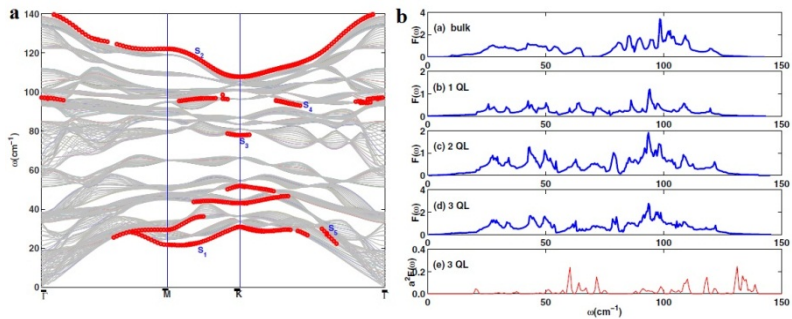

**Figure 12** *ab initio* calculation results from Huang. **a**, The localized surface phonon modes (red circles) of the 3-QLs film. The light gray lines represent the projected bulk phonon dispersions onto the SBZ. **b**, The phonon density of states $F(\omega)$ for $Bi_2Te_3$ thin films with thicknesses of 1-3 QLs as well as for the bulk. The Eliashberg function $\alpha^2 F(\omega)$ for the 3-QLs film is shown in the bottom panel. (Figures are adapted from Ref. [52], reprinted by permission from EPL.)

### 2.4 Other Related Studies

In the sections above, we outlined several studies of EPC on the surface of TIs with the strength of EPC given by the corresponding EPC constant.

There are also numerous related studies concerning the e-p interactions in TIs, which did not give specific numbers for the EPC constant. Some of those studies are reviewed below.

A recent optical infrared study of $Bi_2Se_3$ found the lineshape of the 7.6 meV optical phonon mode is highly asymmetric, which is a signature of strong EPC [53]. This mode is possibly the same mode discovered by HAS measurements experiencing a strong Kohn anomaly [35].

An ultrafast time-resolved differential reflectivity study found three distinct relaxation processes that contribute to the reflectivity changes. The deduced relaxation timescale and the sign of the reflectivity change suggest that EPC and defect-induced charge trapping are the underlying mechanisms for the three processes [54].

A Raman spectroscopy study of few QL $Bi_2Se_3$ nanoplatelets measured the softening and linewidth of several phonon modes. The studies provided evidence that the EPC in the few QL $Bi_2Se_3$ system is remarkably different from that in a bulk sample [55]. This result is consistent with the theory based on the isotropic elastic continuum phonon model [49, 50].

A recent theoretical study showed that, in narrow-gap semiconductors, phonon-induced renormalization of the band gap can culminate in a band inversion at both zero and non-zero temperature. Thus EPC can alter the topological properties of Dirac insulators and semimetals, at both zero and nonzero temperature. This result showed the possible instances in which phonons may lead to the appearance of topological surface states above a crossover temperature in a material that has a topologically trivial ground state [56].

A recent study showed that a terahertz (THz) wave can be generated from the (001) surface of cleaved $Bi_2Se_3$ and Cu-doped $Bi_2Se_3$ single crystals using 800 nm femtosecond pulses [57]. The characteristics of THz emission confirm the existence of a surface optical phonon branch that is renormalized by Dirac fermions in Ref. [35].

## 3. Summary

In summary, the study of EPC on the surface of TIs is of exceptional importance not only for the potential applications for devices based on TIs, but also for the fundamental understanding of the physics behind the interaction of lattice vibrations with massless surface quasiparticles. Thus it is now a vibrant area in both experimental and theoretical condensed matter physics, and many studies are still probing these fascinating materials.

The EPC constant $\lambda$ is, by far, the most convenient and effective parameter to characterize the strength of the interaction of electrons with phonons. Table I summarizes the reported values of the EPC constant $\lambda$ for the TIs from various studies, both theoretical and experimental.

It is obvious that $\lambda$ has a wide range of reported values, from 0 for p-type $Bi_2Te_3$ in one study to 3 for Cu-doped $Bi_2Se_3$ in another study. One should notice that there is disagreement among the reported values even for the same material and the same kind of sample. The inconsistency is mainly due to the following three reasons:

1. First, the complexity of the EPC mechanism makes it hard to obtain a common EPC constant that can be used everywhere. As mentioned in Section 1.2, the EPC can be studied either from the electron or phonon perspective. When determining the EPC constant associated with an electronic state one integrates over all phonon states, and vice versa. Thus one really needs to be cautious when comparing the EPC constants obtained from electron and phonon perspectives.

2. Second, the method of extracting the EPC from experimental data needs to be improved. For example, from ARPES measurements, one can calculate the EPC constant either from the imaginary part of the extracted electron self-energy or from the real part. However, both methods have strict constraints and the obtained values are usually not consistent.

3. Third, although the study of the EPC in TIs should ideally be confined to the surface electronic and phonon states, it is very difficult to exclude effects from bulk states. This is possibly why the thin film and cleaved bulk samples of the same material show very different EPC constants.

**Table I: EPC Constants on the Surface of Topological Insulators**

| Author(s) | Method | Material | Sample | λ | Ref |
|---|---|---|---|---|---|
| R. Hatch et al. | Synchrotron-based ARPES | $Bi_2Se_3$ | Cleaved bulk | 0.25 | [29] |
| Z.-H. Pan et al. | Synchrotron-based ARPES | $Bi_2Se_3$ | Cleaved bulk | 0.08 | [30] |
| C. Chen et al. | Laser-based ARPES | $Bi_2Se_3$ | Cleaved bulk | 0.17 | [34] |
| | | p-type $Bi_2Te_3$ | | 0 | |
| | | n-type $Bi_2Te_3$ | | 0.19 | |
| T. Kondo et al. | Laser-based ARPES | $Bi_2Se_3$ | Cleaved bulk | 3 * | [37] |
| | | $Bi_2Te_3$ | | | |
| | | Cu-doped $Bi_2Se_3$ | | | |
| X. Zhu et al. | HAS | $Bi_2Se_3$ | Cleaved bulk | 0.43 | [47] |
| S. Giraud et al. | Continuum phonon model | $Bi_2Se_3$ | Semi-infinite Bulk | 0.84 | [50] |
| | | | Thin film | 0.42 | |
| | | $Bi_2Te_3$ | Semi-infinite Bulk | 0.13 | [49] |
| G. Q. Huang | *ab initio* calculation | $Bi_2Te_3$ | Slab | 0.05 | [52] |

* This exceptionally large coupling constant contains the contribution from both the EPC and the electron spin-plasmon coupling. See details in Ref.[37]

The results outlined in Table I suggest that more systematic studies of the EPC in TIs need to be undertaken. One possible effective way is to combine the electronic and phonon measurements and study the EPC effect from both phonon and electronic perspectives to finally obtain a consistent result. With the development of instrumentation and improvement of data analysis methods, more detailed and convincing pictures about the EPC on the surface of TIs will surely follow.

**Author Notes**

During the publication process of this article, a lot of new research results about the EPC of topological insulators were published. Some of them are listed in Refs. [58, 59, 60, 61]. Interested readers are strongly recommended to check those new progresses in this field.

**Acknowledgements**

X. Zhu acknowledges support from the International Young Scientist Fellowship from the Institute of Physics, Chinese Academy of Sciences.